\documentclass{Amade}

\usepackage{bm}

\usepackage{amsmath}
\usepackage{amssymb}
\usepackage{graphicx}
\usepackage{subfigure}
\usepackage{verbatim}

\usepackage{pgfbaselayers}
\usepackage{tikz}
\usetikzlibrary{arrows,shapes}

\newcommand*{\norm}[1]{\mathopen\| #1 \mathclose\|}% use instead of $\|x\|$
\def\norm#1{\mathopen\| #1 \mathclose\|}% use instead of $\|x\|$
\def\bx{{\bm x}}
\def\bv{{\bm v}}
\def\bN{{\bm N}}
\def\be{\begin{equation}}
\def\ee{\end{equation}}
\def\bea{\begin{eqnarray}}
\def\eea{\end{eqnarray}}
\def\lb{\label}

\begin{document}

\chapter[Transponder delay
effect in light time calculations]{Transponder delay
effect in light time calculations for deep space navigation}{S. Bertone$^{1,2}$, C. Le Poncin Lafitte$^1$, V. Lainey$^3$ and M.-C. Angonin$^1$}{$^1$ SYRTE - Obs. de Paris - CNRS/UMR8630, UPMC, France\\
$^2$ INAF - Astronomical Observatory of Turin, University of Turin, Italy \\
$^3$ IMCCE - Obs. de Paris - CNRS/UMR8028, UPMC, France \break e-mail: stefano.bertone@obspm.fr}

\authormark{S. BERTONE {et~al.}}

%%%%%%%%%%%%%%%%%%%%%%%%%%%%
\begin{abstract}
%%%%%%%%%%%%%%%%%%%%%%%%%%%%
During the last decade, the precision in the tracking of spacecraft has constantly improved. The discovery of few astrometric anomalies, such as the Pioneer and Earth flyby anomalies, stimulated further analysis of the operative modeling currently adopted in Deep Space Navigation~(DSN). 
Our study shows that some traditional approximations lead to neglect tiny terms that could have consequences in the orbit determination of a probe in specific configurations such as during an Earth flyby. Therefore, we suggest here a way to improve the light time calculation used for probe tracking. 
\end{abstract}

\begin{keys}
space navigation, light time, transponder-based orbit determination \\[7pt]
MSC (2000): 

\end{keys}

%%%%%%%%%%%%%%%%%%%%%%%%%%%%
\section{Introduction}
%%%%%%%%%%%%%%%%%%%%%%%%%%%%
Deep space data processing during the last decade has revealed the presence of anomalies in the form of unexpected accelerations in the trajectory of probes~\cite{2008PhRvL.100i1102A,1998PhRvL..81.2858A}.  The hypothesis made trying to solve this puzzle can be summarized in two main approaches: whether these anomalies are the manifestation of some new physics ~\cite{2008MNRAS.389L..57M,2009PhRvD..79b3505A}, or something is mismodeled in the data processing \cite{2009ScReE2009.7695I,2012PhRvL.108x1101T}. \\

We investigate Moyer's book~\cite{Moyer:2000}, which describes the relativistic framework used by space agencies for data processing.
We know that the ephemeris of a space mission is built from subsequent measures involving the light time of a signal traveling between the Earth and the probe and the solution of the inverse problem. Since the ephemeris is used for both operational (space probe navigation) and scientific goals (measurements for testing fundamental physics), a well defined model is then mandatory for both the interpretation of physical data and the orbit reconstruction.
In this article, we suggest an improvement of the light time modeling focusing on the treatment of the so-called "transponder delay".\\

\noindent This paper is structured as follows.\\
In section 2, we give a brief overview of light time computation as described by the Moyer's book; we show that the transponder's delay (i.e. the time delay between the reception and retransmission of the light signal on board the satellite) is not  accurately taken into account in this model. In section 3 we present an alternative, more precise, modeling. Finally, in section 4, we compare both modelings to highlight their differences and give some conclusions in section 5. \\

Throughout this work we will suppose that space-time is covered by some global barycentric coordinates system $x^{\alpha} = (x^0, \bx)$, with $x^0 = ct$, $c$ being the speed of light in vacuum, $t$ a time coordinate and $\bx = (x^i, i=1,2,3)$. Greek indices run from 0 to 3, and Latin indices from 1 to 3. Here $\bx^b_t$/$\bv^b_t$ represents the position/velocity of body $b$ at time $t$, where $b$ can take the value $GS$ (ground station) or $SC$ (spacecraft). Primed values are related to the {\it Moyer's modeling}, while we will generally use non-primed values for our proposed modeling.

%%%%%%%%%%%%%%%%%%%%%%%%%%%%
\section{Moyer's navigation model}  \lb{sec:Moyermodel}
%%%%%%%%%%%%%%%%%%%%%%%%%%%%
Deep space navigation is based on the exchange of light signals between a probe and at least one observing ground station. The calculation of a coordinate light time, as resumed from~\cite{Moyer:2000}, is quite simple: a clock starts counting as an uplink signal is emitted from ground at $\bx^{GS}_{1'}$. The signal is received by the probe at $\bx^{SC}_{2'}$ and then, after a short delay, reemitted towards the Earth where it is received by a ground station at $\bx^{GS}_{3'}$. The clock stops counting and gives the round-trip light time
\begin{equation}
	\rho' =  \frac{R_{1'2'}}{c} + \frac{R_{2'3'}}{c} +\frac{\Delta (\bx_{1'}^{GS},\bx_{2'}^{SC})}{c} + \frac{\Delta (\bx_{2'}^{SC},\bx_{3'}^{GS})}{c}  +\delta t + \delta C \; , \label{eq:ltJPL}
\end{equation}
where $R_{ab}=\norm{\bx_a-\bx_b}$, $c$ is the speed of light, $\Delta (\bx_{a},\bx_{b})$ is the Shapiro delay~\cite{1993tegp.book.....W}, {while $\delta t$ and $\delta C$ are the transponder delay and other corrections (ex : atmospheric delay ... ) that we will not detail here, respectively.}
The light time $\rho'$ is then used to compute two physical quantities: 
\begin{itemize}
	\item the {\it Ranging}, { related to} the distance between the probe and the ground station can be computed { using}
		\begin{equation}
  			\mathcal R' =  \rho'  -\delta t - \delta C   \label{eq:rangJPL} \; ;
		\end{equation} 
	\item the {\it Doppler}, related to the velocity of the probe with respect to the Earth, is obtained by differentiating two successive light time measurements, $\rho'_s = t_{3s}-t_{1s}$ and $\rho'_e = t_{3e}-t_{1e}$, during a given count interval $T_c = t_{3e}-t_{3s}$. { It has been} showed that 
		\begin{equation}
			\mathcal F' = \frac{\Delta \nu'}{\nu_{1'}} = M_2 f_T (t_1) \frac{ \rho'_e-\rho'_s }{T_c} = M_2 f_T (t_1) \dot \rho' \; ,
		\end{equation}
		 where $M_2$ is a transponder's ratio applied to the downlink signal when it is reemitted towards the Earth and $\dot \rho' = {d \rho'}/{d t}$. 

		Since the {\it Doppler} signal results from the differentiation of the {\it Ranging} signal, all constant or slowly changing terms like $\delta t$ and $\delta C$ obviously cancel out in this modeling.
\end{itemize}

%%%%%%%%%%%%%%%%%%%%%%%%%%%%
\section{Our improved navigation model} \lb{sec:ourmodel}
%%%%%%%%%%%%%%%%%%%%%%%%%%%%
Nevertheless, the electronic delay of some microseconds $\delta t$ due to on board processing of the incoming signal requires to consider a different position of the spacecraft at reemission time. { In the following, we study} its consequences on light time modeling for \textit{Ranging} and \textit{Doppler} calculations.\\ 

For this purpose, we introduce an improved light time model $\rho$ taking into account four events (one more with respect to Moyer's model): the emission from the ground station at $\bx^{GS}_1$, the reception by the probe at $\bx^{SC}_2$, the reemission at $\bx^{SC}_3$ and the reception at ground at $\bx^{GS}_4$. The additional event $\bx^{SC}_3=\bx^{SC}_{2+\delta t}$ accounts for this small delay of $\delta t$$\approx$$2.5$ $\mu s$ (at least for modern spacecraft) so that we get
\begin{equation} \lb{eq:ltSYRTE}
	\rho =  \frac{R_{12}}{c} + \frac{R_{34}}{c} +\frac{\Delta (\bx_{1}^{GS},\bx_{2}^{SC})}{c} + \frac{\Delta (\bx_{3}^{SC},\bx_{4}^{GS})}{c} +\delta t + \delta C \; .
\end{equation}
{ Similarly to Section~\ref{sec:Moyermodel}, we then} use $\rho$ to compute {\it Ranging} $\mathcal R$ and {\it Doppler} $\mathcal F$ observables as  
\begin{subequations}
\begin{eqnarray}
	\mathcal R &=&  \rho  -\delta t - \delta C, \label{eq:rangSYR} \\
	\mathcal F &=& \frac{\Delta \nu}{\nu_{1}} = M_2 f_T (t_1) \dot \rho \; , 
\end{eqnarray}
\end{subequations}
{where $\dot \rho = \left( \rho_e-\rho_s \right )/T_c$.} 
{ In principle, we have that} $\mathcal R - \mathcal R' \ne 0$ and $\mathcal F - \mathcal F' \ne 0$, since primed and non-primed events are \emph{a priori} separated.

%%%%%%%%%%%%%%%%%%%%%%%%%%%%
\section{Comparison of the two modelings}
%%%%%%%%%%%%%%%%%%%%%%%%%%%%
{ To compare the two modelings presented in sections~\ref{sec:Moyermodel} and~\ref{sec:ourmodel}, we shall define { the difference between the computed light times}
\be \lb{eq:deltarho}
	\Delta \rho= \rho - \rho' = t_{1'} - t_{1} \; ,
\ee
{ where we use} $t_{3'}=t_4$ and $t_{2'}=t_3$. Let us then} analyze the supplementary event $\bx^{SC}_3=\bx^{SC}_{2+\delta t}$. This term is implicitly related to $\delta t$ by the first order development 
\be \lb{eq:xSC3}
	\bx^{SC}_3=\bx^{SC}_2+\delta t \; \bv^{SC}_2+ O (\delta t^2) \; ,
\ee 
{ which is usually neglected in the standard light time modeling.}
%In practice, this $\delta t_{23}$ delay is calibrated by space agencies : it is added to $\rho$ when computing $\mathcal R$ and seems to have no consequence on the differential \textit{Doppler} $\mathcal F$ since it's a constant. 
%However, as $\delta t_{23}$ also appears implicitly in the expression of $\bx^{SC}_3$, this is not true. 

{ The implications of this mismodeling are given by
\begin{equation}
  \Delta \rho = \rho - \rho' =  \frac{\delta t}{c} \frac{ \left( \bv^{SC}_2\cdot \bN_{12} \right) } { 1 +  \frac {1}{c} \left(\bv^{GS}_1 \cdot \bN_{12} \right) }
\, \text{ with } \, \bN_{12} \equiv \frac{\bx^{SC}_2-\bx^{GS}_1}{\norm{\bx^{SC}_2-\bx^{GS}_1}} \; , \label{eqn:delta_rho_def}
\end{equation} 
{ where we used Eq.~\eqref{eq:ltJPL}, Eq.~\eqref{eq:ltSYRTE} and Eq.~\eqref{eq:xSC3} into Eq.~\eqref{eq:deltarho} and defining $\bN_{12}$ as} the Minkowskian direction between the ground station and the probe.\\

Equation~\eqref{eqn:delta_rho_def}} highlights the presence of an extra non-constant term, directly proportional to the transponder delay and neglected in Moyer's model. This term also depends on the position and velocity of both the probe and the ground station. Neglecting it would actually lead to a wrong determination of the epoch $t_1$ and to an error in both \textit{Ranging} and \textit{Doppler}. 
%In fact, extracting the downlink signal, we get $|\bx^{SC}_{2'}-\bx^{GS}_{1'}| \neq |\bx^{SC}_{2}-\bx^{GS}_{1}|$ since $t_2\neq t_2'$ and $t_1\neq t_1'$.

%%%%%%%%%%%%%%%%%%%%%%%%%%%%
\section{Application to real spacecraft orbits}
%%%%%%%%%%%%%%%%%%%%%%%%%%%%
In order to evaluate the magnitude of the additional term in Eq.~\eqref{eqn:delta_rho_def}, we computed $\Delta \rho$ (giving the difference between the  {\it Ranging} calculated with the two models) and $\Delta \dot \rho =  \dot\rho - \dot\rho'$ (related to the difference of the {\it Doppler} calculated by the two models) for the observation of a probe. We used the real orbit of some probes (Rosetta, NEAR, Cassini, Galileo) during their Earth flyby, which is a particularly favorable configuration. We used the NAIF/SPICE toolkit~\cite{2011epsc.conf...32A} to retrieve the ephemeris for probes and planets to be used in the computation.

\begin{figure}[ht] 
	\centering
	\includegraphics[width=0.78\textwidth]{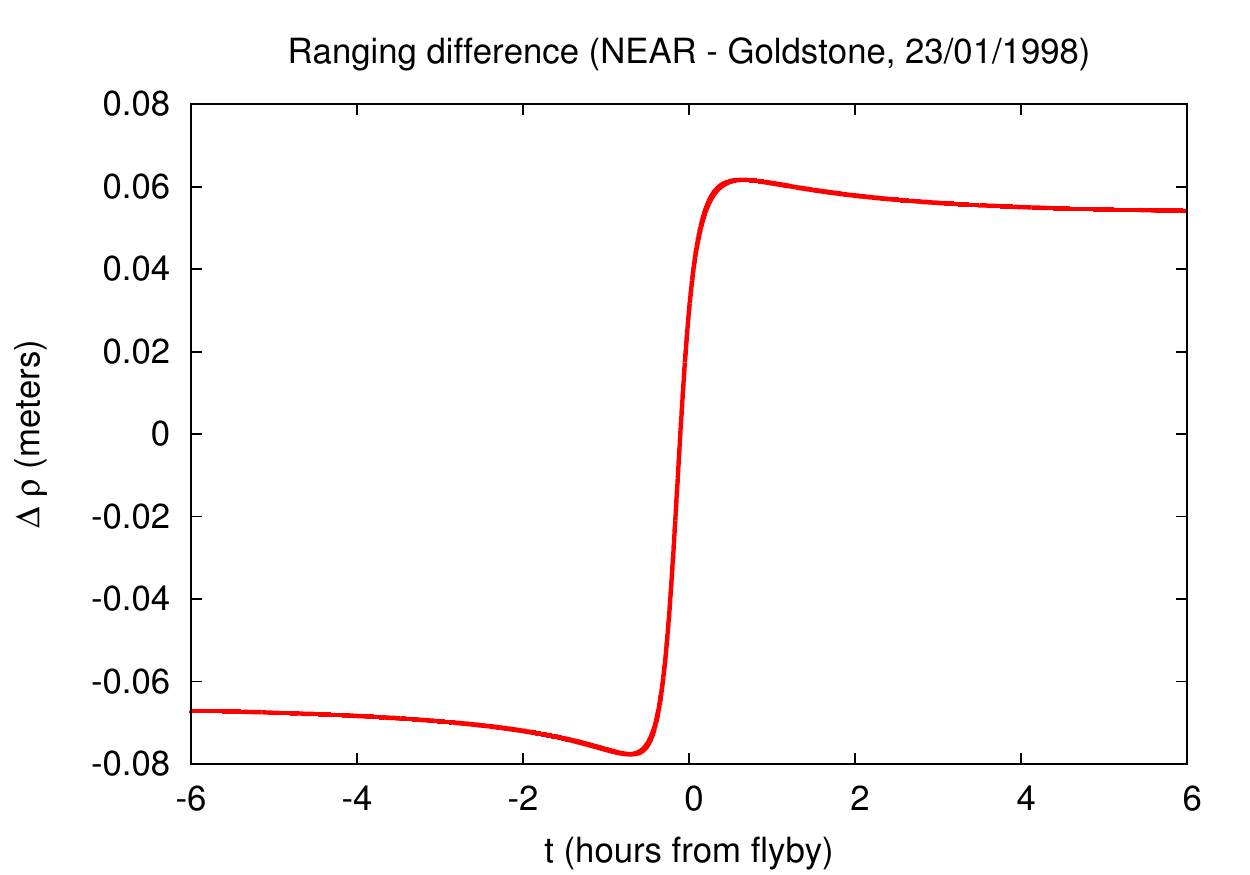}
	\caption{{\it Ranging} difference $\Delta \rho$ (meters - hours from flyby) during NEAR Earth flyby.}
	\label{fig:range}
\end{figure}

\begin{figure}[ht] 
	\centering
	\includegraphics[width=0.78\textwidth]{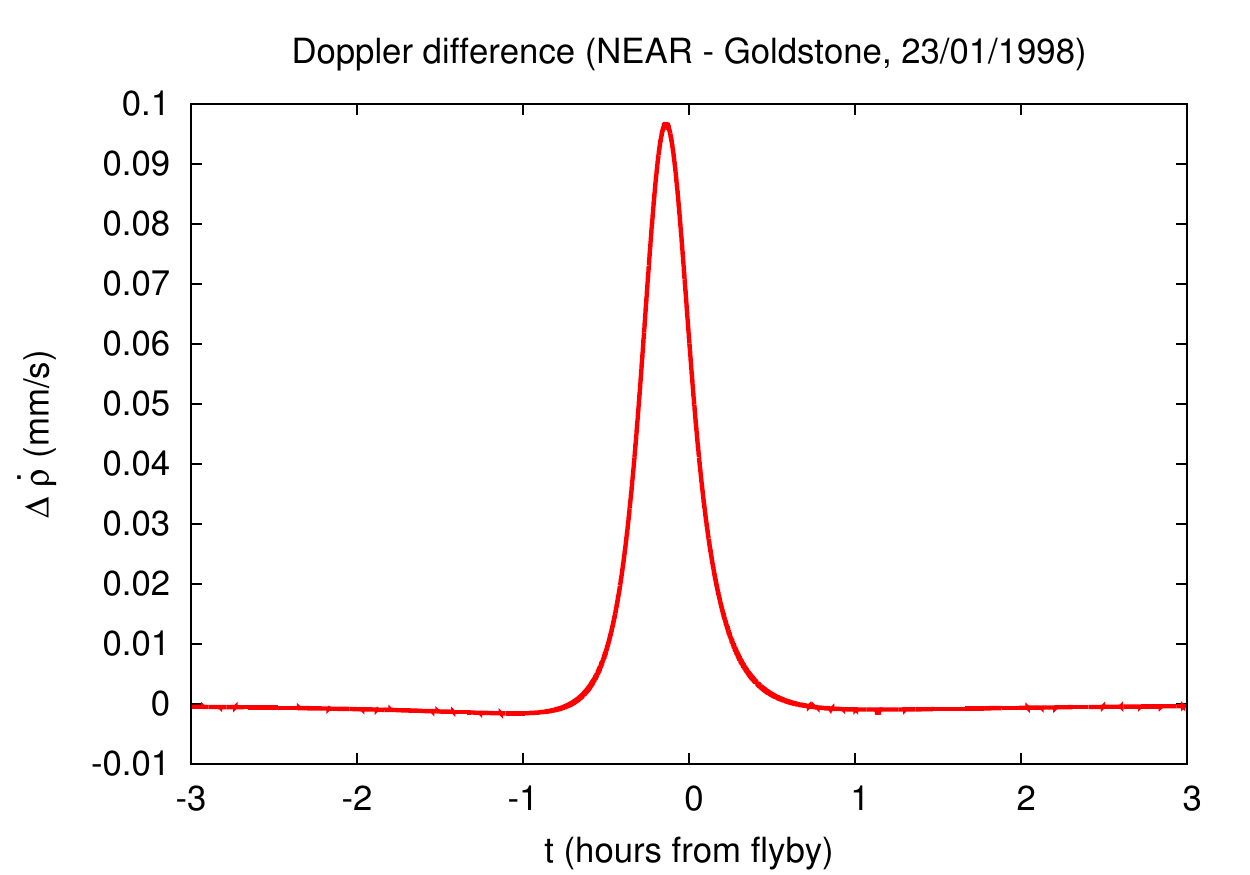}
	\caption{{\it Doppler} difference $\Delta \dot \rho$ (mm/s - hours from flyby) during NEAR Earth flyby.}
	\label{fig:dop}
\end{figure}

Computing Eq.~\eqref{eqn:delta_rho_def} and it's time derivative for the NEAR probe during its Earth flyby on the 23 January 1998, we found a difference of the order of some $cm$ for the probe distance $c \Delta \rho$ calculated by the two models and a difference up { to several $10^{-2} \; mm/s$} at the instant of maximum approach { for its velocity}. These results are shown in Figures~\ref{fig:range} and\ref{fig:dop}.\\

\begin{figure}[ht]
	\centering
	\includegraphics[width=0.78\linewidth]{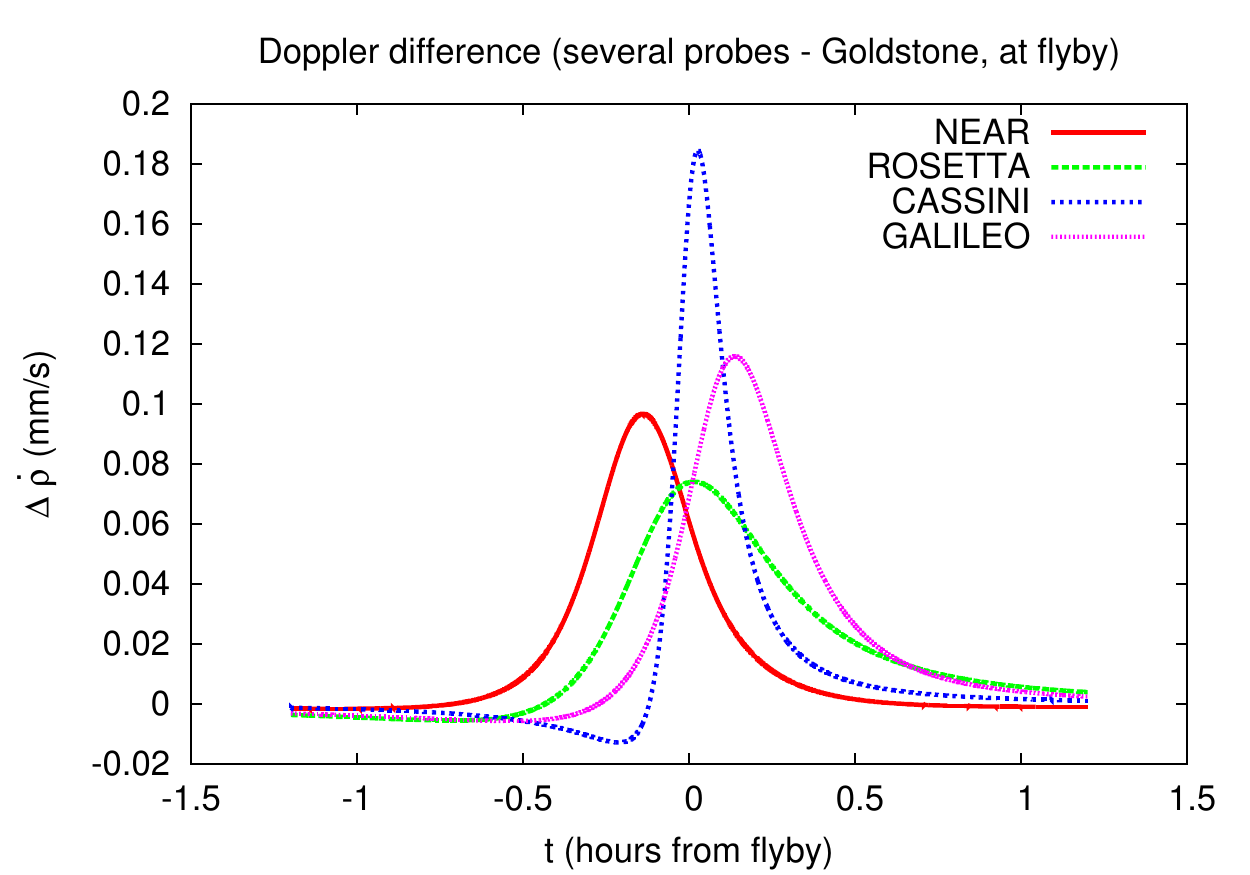}
	\caption{{\it Doppler} difference $\Delta \dot \rho$ (mm/s - hours from Earth flyby) for several probes with respect to Goldston DSN station. The results highlight the high variability of the effect on Doppler measurements.}
	\label{fig:dop1}
\end{figure}

In order to highlight the high variability of the transponder delay effect on Doppler { measurements}, we computed $\Delta \dot \rho$ for different probes in different configurations with respect to the observing station. The results are exposed in Figure~\ref{fig:dop1} and show that this delay cannot be simply calibrated at the level of light time calculation nor neglected in the {\it Doppler} calculation.

\clearpage
%%%%%%%%%%%%%%%%%%%%%%%%%%%%
\section{Conclusions}
%%%%%%%%%%%%%%%%%%%%%%%%%%%%
It seems obvious from our results that the influence of the transponder delay cannot be reduced to a simple calibration without taking some precautions. It is { indeed} responsible for a tiny effect on the computation of light time and has an impact on both \textit{Ranging} and \textit {Doppler} determination. We represent it by a more complete modeling, considering four epochs instead of three. In order to test the amplitude and variability of this effect on real data, we compute its influence on some real probe-ground station configurations during recent Earth flybys (NEAR, Rosetta, Cassini and Galileo). \\
%and we computed the alternative orbit fitted to \textit{Ranging} and \textit{Doppler} data from our light time model. 

The observables calculated using Moyer's model and our improved model show differences of the order of several $cm$ and of $0.1 \; mm/s$ for the \textit{Ranging} and the \textit {Doppler}, respectively.
Such an error is acceptable for most operational goals at present. Anyway, we { shall} highlight that this error is directly proportional to the transponder delay and that for past missions, whose data are still largely used for scientific purposes, transponders were more than $10^3$ times slower that today. In the future too, increasing ephemeris precision~\cite{2014AcAau..94..699I} should be followed by the development of faster transponders or by the use of a more precise model. 

\vspace*{0.2cm}

\textit{Acknowledgements.} The authors are grateful to the anonymous referees for their detailed review, which allowed to improve the paper. S. Bertone and C. Le Poncin-Lafitte are grateful to the financial support of CNRS/GRAM .

\end{document}